\documentclass[sigconf]{acmart}

\usepackage{array}

\AtBeginDocument{%
  }

\copyrightyear{2025}
\acmYear{2025}
\setcopyright{cc}
\setcctype{by}
\acmConference[SIGIR '25]{Proceedings of the 48th International ACM SIGIR Conference on Research and Development in Information Retrieval}{July 13--18, 2025}{Padua, Italy}
\acmBooktitle{Proceedings of the 48th International ACM SIGIR Conference on Research and Development in Information Retrieval (SIGIR '25), July 13--18, 2025, Padua, Italy}\acmDOI{10.1145/3726302.3730213}
\acmISBN{979-8-4007-1592-1/2025/07}
\begin{document}

%%
%% The "title" command has an optional parameter,
%% allowing the author to define a "short title" to be used in page headers.
\title{In a Few Words: Comparing Weak Supervision and LLMs for Short Query Intent Classification}

%%
%% The "author" command and its associated commands are used to define
%% the authors and their affiliations.
%% Of note is the shared affiliation of the first two authors, and the
%% "authornote" and "authornotemark" commands
%% used to denote shared contribution to the research.

\author{Daria Alexander}
\affiliation{%
  \institution{Radboud University}
  \city{Nijmegen}
  \country{the Netherlands}}
\email{daria.alexander@ru.nl}

\author{Arjen P. de Vries}
\affiliation{%
  \institution{Radboud University}
  \city{Nijmegen}
  \country{the Netherlands}}
\email{arjen.devries@ru.nl}

%%
%% By default, the full list of authors will be used in the page
%% headers. Often, this list is too long, and will overlap
%% other information printed in the page headers. This command allows
%% the author to define a more concise list
%% of authors' names for this purpose.
\renewcommand{\shortauthors}{Daria Alexander and Arjen P. de Vries}
%% No italics, no superscripts
%% Use footnote or author note to identify equal contribution and/or contact author info

%%
%% The abstract is a short summary of the work to be presented in the
%% article.
\begin{abstract}
User intent classification is an important task in information retrieval. Previously, user intents were classified manually and automatically; the latter helped to avoid hand labelling of large datasets. Recent studies explored whether LLMs can reliably determine user intent. However, researchers have recognized the limitations of using generative LLMs for classification tasks. In this study, we empirically compare user intent classification into informational, navigational, and transactional categories, using weak supervision and LLMs. Specifically, we evaluate LLaMA-3.1-8B-Instruct and LLaMA-3.1-70B-Instruct for in-context learning and LLaMA-3.1-8B-Instruct for fine-tuning, comparing their performance to an established baseline classifier trained using weak supervision (ORCAS-I). Our results indicate that while LLMs outperform weak supervision in recall, they continue to struggle with precision, which shows the need for improved methods to balance both metrics effectively.\end{abstract}

%%
%% The code below is generated by the tool at http://dl.acm.org/ccs.cfm.
%% Please copy and paste the code instead of the example below.
%%


%%
%% Keywords. The author(s) should pick words that accurately describe
%% the work being presented. Separate the keywords with commas.
%% A "teaser" image appears between the author and affiliation
%% information and the body of the document, and typically spans the
%% page.

%%
%% This command processes the author and affiliation and title
%% information and builds the first part of the formatted document.
\maketitle

\section{Introduction}

Large language models (LLMs) have gained increasing attention for generative tasks such as question answering, dialogue, and summarization \cite{tan2023can,jin2024comprehensive,li2024pre}. Researchers have also investigated their potential for discriminative tasks, including user intent classification \cite{yadav2024pag,dubiel2024device}. If LLMs could distinguish user intent well, this would enable simpler system architectures, delegating more tasks of the retrieval system to the LLM component.

However, recent research points out that LLMs are in their core designed as generative models, and therefore prioritize their generative capabilities, often at the expense of discriminative capabilities \cite{sun2023llm}. Many of the reasoning rationales cannot effectively support the claims or address the reasoning required, indicating that the model often does not truly understand the content and reasoning behind the question, even if it arrives at the correct answer \cite{xu2024llms}. This finding may also apply to user intent classification, where models have been found to struggle with making correct predictions, and, even worse, may generate out-of-vocabulary intent labels \cite{yadav2024pag}.

To annotate queries with user intent, previous studies used manual annotation \cite{broder2002taxonomy, rose2004understanding, russell2009task}, combining manual and automatic annotation  \cite{figueroa2015exploring, bolotova2022non}, or fully automatic annotation \cite{jansen2007web, mohasseb2019customised, alexander2022orcas}. Weak supervision is a popular machine learning approach that allows the annotation of large amounts of data at a low cost. Instead of (or along with) gold labeled data, weak supervision uses noisy, limited, or imprecise sources. The approach is straightforward to implement, but relies on rules and heuristics, without semantic understanding.

We assess whether LLMs can outperform state-of-the-art approaches for short query intent classification that are based on weak supervision. Our classification is based on the intent taxonomy established by Broder \cite{broder2002taxonomy}, which divides user intent into three categories: informational, navigational and transactional. An informational intent refers to acquiring information from a website, a navigational intent seeks to reach a particular website, and a transactional intent refers to obtaining services from a website.

We compare results using LLMs to those using the ORCAS-I weak supervision classifier \cite{alexander2022orcas}, which annotated the queries in the ORCAS dataset into informational, navigational, and transactional intent. A key advantage of utilizing the ORCAS-I dataset is its scale, with over 10 million queries from Bing search logs. Also, it does not suffer from the privacy issues of the Dogpile transaction log \cite{jansen2007web} and the AOL web query collection \cite{pass2006picture}, used in previous studies for intent classification.

Given that ORCAS-I queries are both short and representative of general search behavior, it would be interesting to explore whether LLMs can derive more meaningful information from a large number of short queries, especially in the context of fine-tuning. To investigate this, we perform in-context learning (ICL) and weakly supervised fine-tuning (wSFT) experiments. We utilize LLaMA-3.1-8B-Instruct and LLaMA-3.1-70B-Instruct for ICL, and, considering resource constraints, only LLaMA-3.1-8B-Instruct for wSFT. We compare our results with the results of the ORCAS-I weak supervision classifier that was used to annotate the ORCAS-I datasets.

Our research questions are the following: 

\begin{itemize}
    \item Can LLMs classification outperform weak supervision classification for a large set of short queries?
    \item What LLM approaches show the best performance?
\end{itemize}

Our work makes the following contributions:
\begin{itemize}
    \item It highlights the challenges that LLMs encounter when performing user intent classification on short queries.
    \item It explores how weak supervision can be effectively used alongside LLMs.
\end{itemize}

For reproducibility and transparency, we make our code for in-context learning and fine-tuning scripts publicly available\footnote{See \url{https://github.com/daria87/intent_classification_llms}}.

\section{Related work}

\subsection{Automatic user intent classification}

Automatic classification of user intent started with approaches that used keywords and heuristics. Jansen et al.\ \cite{jansen2008determining} established a rule-based approach for automatic classification of queries into informational, navigational and transactional intents. They defined query characteristics related to query length, to the occurrence of specific words, and to information about the search session. Other studies utilized rule-based approaches as a first step preceding machine learning, such as performing k-means clustering \cite{kathuria2010classifying, chandrakala2024intent}, natural language processing (NLP) techniques like POS-tagging \cite{figueroa2015exploring, alexander2022orcas}, named entity recognition and dependency parsing \cite{figueroa2015exploring}. 

Yadav et al.\ \cite{yadav2024pag} applied LLMs to intent prediction for paraphrasing queries, and showed promising results with a substantial reduction in classification error. Zhang et al.\ \cite{zhang2024clamber} utilized LLMs for query intent disambiguation, but found difficulties with ambiguity identification and clarification. This was further elaborated by Liu et al.\ \cite{liu2023we}, who found GPT-4 disambiguation to be correct only 30\% of the time, on a dataset annotated with ambiguity levels.

\subsection{Using LLMs for classification tasks}

Text classification is a task that aims to assign predefined labels (e.g., sentiment, topic, etc.) to a given text. With the introduction of LLMs, researchers started using in-context learning for classification tasks. Han et al.\ \cite{han2022prompt} designed sub-prompts and applied logic rules to compose sub-prompts into final prompts. Liu et al.\ \cite{liu2021makes} retrieved in-context examples according to their semantic similarity to the test samples. Sun et al.\ \cite{sun2023llm} proposed a strategy that involves getting clues first (such as keywords, semantic clues), then perform reasoning and then make a decision about the correct class. 

However, LLMs suffer from `hallucination', which can impact the classification task. Xu et al.\ \cite{xu2024llms} excluded golden labels from labeling space and found that LLMs still attempt to select from the available label candidates; even when none are correct. \cite{rouzegar2024enhancing} therefore suggested a hybrid approach that combines GPT-3.5 with human annotations, hoping to improve model performance at lower data annotation expenses. Liu and Shi \cite{liu2024poliprompt} selected a small pool of diverse examples to represent the broader dataset, guiding the LLMs to extract classification rules and write enhanced prompts based on accurate human-annotated texts.

\section{ORCAS-I intent classifier}

The ORCAS-I intent classifier \cite{alexander2022orcas} was initially created to classify general search queries. It was used to annotate ORCAS dataset that was aggregated based on a subsample of Bing’s 26-month logs to January 2020. This dataset contains 10.4 million distinct queries and 18.8 million clicked query-URL pairs \cite{craswell2020orcas}. As these are general search queries, they are short, with a mean length of 3.25 words \cite{alexander2022orcas}. The dataset is released with 2M random sample for pretraining: ORCAS-I-2M. The authors also released an ORCAS-I-gold test set: 1000 randomly selected queries from the original ORCAS dataset that were not in the ORCAS-I-2M dataset. These 1000 queries were annotated manually by IR experts.

The ORCAS-I intent classifier is based on weak supervision. Queries are annotated with Snorkel labeling functions using keywords, heuristics and NLP techniques like POS-tagging. Majority voting determines the final classification results. For some intent categories, using the clicked URLs provided in the ORCAS dataset helps to improve the results.

The classifier contains two levels of labeling functions according to the taxonomy suggested by the authors. On the first level, the queries are classified into informational, navigational and transactional intent. On the second level, informational queries are further classified into factual and instrumental intent. Informational queries that are neither factual nor instrumental are classified as abstain. 

\section{Experimental setup}

We compare the query-only weak supervision results using ORCAS-I with LLM based intent classification, using both ICL and wSFT. We limit ourselves to the first level of the taxonomy suggested by \cite{alexander2022orcas}, classifying the queries into informational, navigational and transactional categories, because classifying informational queries into the abstain category at the second level would be difficult using ICL. We do not use the clicked URLs, to avoid reliance on information that becomes available only after the user has submitted a query and has clicked on specific pages. We evaluate our approaches on the ORCAS-I-gold test set.

\subsection{In-context learning (ICL)}
For ICL, we experimented with LLaMA-3.1-8B-Instruct and LLaMA-3.1-70B-Instruct on the ORCAS-I-gold test set, as the version 3.1 was considered stable by researchers \cite{morishita2025enhancing,hemied2024re}, and the Instruct models were already optimized for instruction-following \cite{dubey2024llama}. For inference we access LLaMA models through deepinfra API\footnote{https://deepinfra.com/}.

\begin{figure}
  \centering
  \includegraphics[width=0.4\textwidth]{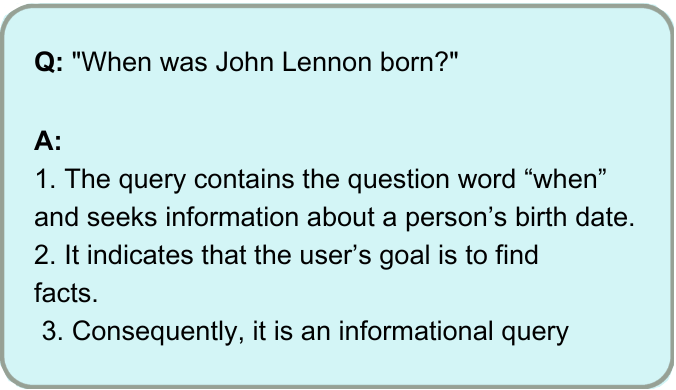} % \columnwidth ensures one-column fit
  \caption{An example of clue-and-reasoning provided to LLaMA-3.1-8B-Instruct and LLaMA-3.1-70B-Instruct.}
  \label{fig:clue_and_reas_eg}
\end{figure}

We used four different scenarios for in-context learning:
\begin{itemize}
    \item definintions of user intent categories only
    \item definitions + keywords
    \item definitions + keywords + few-shot examples
    \item definitions + keywords + few-shot examples + clue-and-reasoning prompting
\end{itemize}
With each subsequent scenario, we increase the amount of information given to the model. When adding keywords, we specify keywords that characterize the intent, such as ``download'' and ``buy'' for transactional and ``login'' and ``site'' for navigational intent. For few-shot prompting, popularized by Brown et al.\ \cite{brown2020language}, we give the model 5 examples per category, along with their correct classes. For clue-and-reasoning prompting (Sun et al.\ \cite{sun2023llm}), we provide each example with an explanation of the classification process. This approach adapts chain-of-thought prompting \cite{wei2022chain} to classification tasks and involves 1.\ clue collection (such as keywords, contextual information, semantic meaning), 2.\ reasoning and 3.\ decision making. Figure \ref{fig:clue_and_reas_eg} shows an example for informational intent.

\subsection{Weakly Supervised Fine-Tuning (wSFT)}

Fine-tuning an LLM can increase its effectiveness beyond that achieved with ICL \cite{liu2022few, mosbach2023few}. Our weakly supervised fine-tuning experiments are based on LLaMA-3.1-8B-Instruct, downloaded locally from the official Meta checkpoint. We use random samples of different sizes, drawn from the ORCAS-I-2M dataset, to determine how the number of examples would affect the performance of the model.

\textbf{Sampling:} We fine-tuned with stratified random samples of sizes 3K, 6K, 15K, 30K, 45K and 60K, ensuring that each user intent category (informational, navigational, and transactional) contains an equal number of examples (i.e., a 45K data sample would comprise 15K informational, 15K navigational and 15K transactional queries). The samples were tokenized using the LLaMA-3.1-8B-Instruct tokenizer, with queries truncated to 32 tokens to fit model constraints (affecting between 0\% and 0.01\% of queries).

After observing no more improvements beyond 45K examples, we expanded the sample using additional queries that are pre-selected for a high confidence in their label predictions. We ran inference using the best model so far (8B-Instruct fine-tuned on 45K) on a 50K query sample, assigning confidence levels to the predictions. From that sample we randomly choose equal numbers of 5K informational, 5K navigational and 5K transactional queries, with high confidence intent predictions. This selection process was repeated for different thresholds of $0.88$, $0.90$, $0.95$ and $0.97$. To ensure that wSFT covered a diverse range of queries, we supplemented the 15K high-confidence queries with 45K additional randomly sampled queries (excluding any overlap), resulting in a total of 60K queries per threshold.

\textbf{LoRA:} For faster fine-tuning we used low-rank adaptation (LoRa). We only fine-tuned the LoRA adapter weights and kept the base model frozen. After training, LoRA adapters were saved separately and later merged with the base model for inference. For each training set size, we reserved 80\% of the data for training and 20\%  for validation to monitor the model's performance. The development set was randomly sampled from the training data, ensuring a representative distribution across all intent categories.

\textbf{Hyperparameters:} We fine-tuned LLaMA-3.1-8B-Instruct on a NVIDIA A100 40GB GPU for 7 epochs, with learning rate=2e-5, batchsize=8, and gradient accumulation step=8. When accuracy on the validation set stagnated after 30K training examples, we decreased the learning rate to 1.5e-5 and increased batch size to 16. To optimize fine-tuning we used dynamic hyperparameter tuning, decaying learning rate adaptively, increasing LoRA rank when performance remained low, and adjusting weight decay to diminish overfitting. Early stopping was used when no improvement was observed.

\section{Results}
We compare the results of our experiments to the results of ORCAS-I intent classifier on ORCAS-I-gold test. As we are interested in the earliest stage of retrieval, we compare our results to the results obtained using query-only retrieval.

\subsection{In-context learning results}

\begin{table}[bth]
    \centering
    \caption{Macro average scores comparison for in-context learning using LLaMA-3.1-8B-Instruct and LLaMA-3.1-70B-Instruct. Statistically significant difference from the baseline is determined by a paired permutation test (5000 iterations) and Bonferroni correction ($p < 0.05$).}
    \label{tab:icl_performance}
    \resizebox{\columnwidth}{!}{%
    \begin{tabular}{
        >{\raggedright\arraybackslash}p{7cm}
        >{\raggedright\arraybackslash}p{1.2cm}
        >{\raggedright\arraybackslash}p{1.2cm}
        >{\raggedright\arraybackslash}p{1.2cm}
    }
        \hline
        \textbf{Model} & \textbf{Precision} & \textbf{Recall} & \textbf{F1-score} \\
        \hline
        \hline
        \textbf{ORCAS-I intent classifier} & \textbf{0.865} & 0.701 & \textbf{0.743} \\
        \hline
        LLaMA-3.1-8B definition only & 0.378\textsuperscript{↓*} & 0.379\textsuperscript{↓*} & 0.377\textsuperscript{↓*} \\
        LLaMA-3.1-8B definition + keywords & 0.467\textsuperscript{↓*} & 0.443\textsuperscript{↓*} & 0.453\textsuperscript{↓*} \\
        LLaMA-3.1-8B definition + keywords + few-shot & \underline{0.573}\textsuperscript{↓*} & \underline{0.584}\textsuperscript{↓*} & \underline{0.577}\textsuperscript{↓*} \\
        LLaMA-3.1-8B clue-and-reasoning & 0.545\textsuperscript{↓*} & 0.530\textsuperscript{↓*} & 0.529\textsuperscript{↓*} \\
        \hline
        LLaMA-3.1-70B definition only & 0.682\textsuperscript{↓*} & 0.577\textsuperscript{↓*} & 0.610\textsuperscript{↓*} \\
        LLaMA-3.1-70B definition + keywords & \underline{0.692}\textsuperscript{↓*} & 0.703 & \underline{0.693} \\
        LLaMA-3.1-70B definition + keywords + few-shot & 0.605\textsuperscript{↓*} & \underline{\textbf{0.737}} & 0.631\textsuperscript{↓*} \\
        LLaMA-3.1-70B clue-and-reasoning & 0.676\textsuperscript{↓*} & 0.721 & 0.667 \\
        \hline
    \end{tabular}%
    }
\end{table}

The results in Table~\ref{tab:icl_performance} show that, as expected, 70-B-Instruct shows much better results than 8B-Instruct. For 8B-Instruct the best prompting strategy was to use definition, keywords and few-shot examples. Interestingly, for 70B-Instruct the best strategy was definition and keywords, although recall improved for definition, keywords and few-shot examples. This might mean that a model with a bigger number of parameters already gets enough information for the classification of short queries without being provided with additional examples.

\begin{figure}
  \centering
  \includegraphics[width=0.4\textwidth]{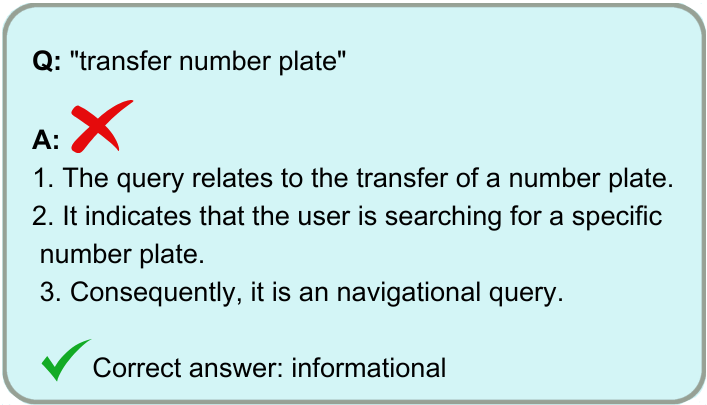} % \columnwidth ensures one-column fit
  \caption{An example of misclassification by LLaMA-3.1-70B-Instruct when using clue-and-reasoning prompt.}
  \label{fig:clue_and_reas_error}
\end{figure}

The clue-and-reasoning strategy does not provide superior results for either model, which means that adding reasoning in the prompt for user intent classification may generate incorrect intermediate steps, an example shown in Figure ~\ref{fig:clue_and_reas_error}. Overall, LLaMA-70B-Instruct outperforms the ORCAS-I intent classifier recall-wise but this difference is not statistically significant. However, both LLMs perform significantly worse than ORCAS-I precision-wise. 

\subsection{Fine-tuning results}

\subsubsection{Random sampling}

\begin{figure}
  \centering
  \includegraphics[width=0.45\textwidth]{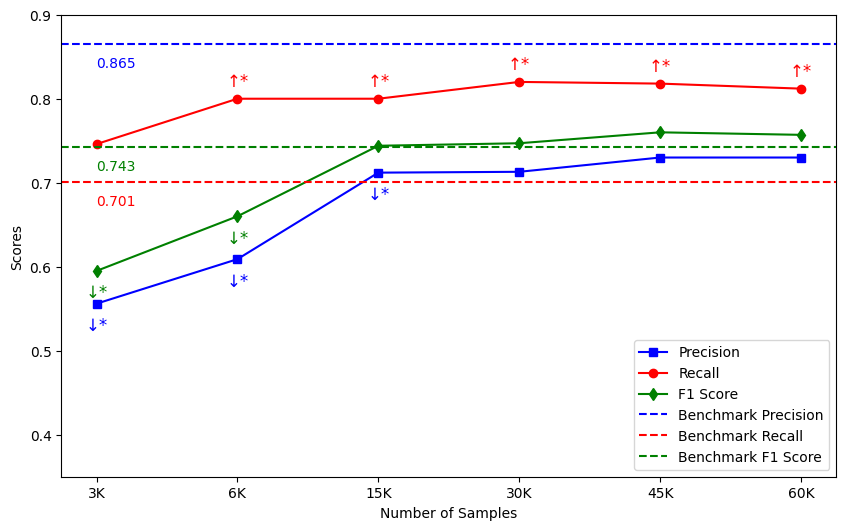} % \columnwidth ensures one-column fit
  \caption{Macro average scores for fine-tuning LLaMA-3.1-8B-Instruct on different random sample sizes, compared to ORCAS-I classifier macro average precision, recall and F1 score on the ORCAS-I-gold test set. Statistically significant difference from the baseline is determined by a paired permutation test (5000 iterations) and Bonferroni correction ($p < 0.05$).}
  \label{fig:finetuning_results}
\end{figure}

As shown in Figure~\ref{fig:finetuning_results}, precision, recall, and F1 score improve rapidly up to 15K samples, after which the improvement rate slows down, with diminishing returns beyond 30K. It indicates that additional data may have limited impact on further optimization. Beyond 45K examples the performance gain drops slightly, which shows that the model starts learning from weak supervision noise. 

Overall, fine-tuned LLaMA-8B-Instruct significantly outperforms the ORCAS-I classifier for recall (except for 3K examples) but continues to struggle with precision, which is similar to its behavior for ICL. An explanation may be found in the architecture of LLMs, which are designed to generate broad, associative responses by capturing different linguistic patterns, favoring inclusiveness over strict relevance; but diverting from the initial query's intent.

\subsubsection{Experiments with confidence thresholds}

\begin{table}[ht]
    \centering
    \caption{Comparison of macro average scores for LLaMA-3.1-8B-Instruct fine-tuned on 60K examples under different settings: a fully random sample and a 45K random sample + 15K high-confidence queries (thresholds 0.88, 0.90, 0.95, 0.97).}
    \label{tab:add_confidence}
    \resizebox{\columnwidth}{!}{%
    \begin{tabular}{lccc}
        \hline
        \textbf{Model} & \textbf{Precision} & \textbf{Recall} & \textbf{F1-score} \\
        \hline
        \hline
        LLaMA-3.1-8B 60K random & 0.730 &  0.812  & 0.757 \\
        \hline
        LLaMA-3.1-8B 45K random + 15K hc. (th. 0.88)& \textbf{0.743} & 0.813 &  \textbf{0.769}\\
        LLaMA-3.1-8B 45K random + 15K hc. (th. 0.90)&  0.740& 0.811& 0.765\\
        LLaMA-3.1-8B 45K random + 15K hc. (th. 0.95) & 0.729&  \textbf{0.814} & 0.761  \\
        LLaMA-3.1-8B 45K random + 15K hc. (th. 0.97) & 0.736 & 0.806 & 0.759 \\
        \hline
    \end{tabular}%
    }
\end{table}

Table ~\ref{tab:add_confidence} shows that adding queries with high confidence labels improved precision over 60K random examples baseline, with the threshold of 0.88 having the best results. At higher thresholds precision drops, which indicates over-filtering. The results suggest that setting the threshold too high can eliminate some important information and negatively impact fine-tuning performance. 

Despite an increase of precision, recall remains relatively stable for high confidence queries, which indicates that adding these helps to reach a better balance between precision and recall. As short queries lack context, filtering them based on high-confidence labels helps the model to better learn the distinctions between different intent categories.

\section{Conclusion}
We set out to compare weak supervision and LLMs for classifying short queries into informational, navigational and transactional intent. We find that for this task, weakly supervised fine-tuning gives better results than in-context learning. Fine-tuning LLaMA-8B-Instruct with a sample of 45K randomly selected queries from ORCAS-I, expanded with 15K high confidence queries (at a threshold of 0.88), leads to the best classification results when using LLMs.

Annotating short queries (mean length of 3.25 words in ORCAS-I dataset) with user intent raises several challenges. The intent of those queries is often ambiguous. Approaches such as paraphrasing or query expansion could potentially help the LLMs to better grasp the intent, although the lack of context might make the model shift away from the correct intent.

An important factor to consider is the cost of training. Our best model, fine-tuned on 60K examples, takes $\sim 10$ hours for fine-tuning and another 1 hour 40 minutes for the inference on 50K queries to select 15K high-confidence samples; and 2 minutes for inference on the test set. In total, just under 12 hours, on a NVIDIA A100 40GB GPU. Alternatively, training the label model and aggregating labels for 60K examples by ORCAS-I classifier takes 10 minutes on CPU; for the test set it takes $\sim 10$ seconds. Clearly, the ORCAS-I classifier currently offers the better solution, both in cost and effectiveness.

The LLMs achieve higher recall, indicating that they may indeed leverage semantic understanding from the queries rather than `just' heuristics or syntactical patterns. However, unless we manage to further enhance precision, this does not lead to improved intent classification in an operational system. The LLMs' high recall capabilities could still be interesting to improve the intent classification of short queries. One approach to consider creates a pipeline with the LLM in front of the ORCAS-I classifier, where the LLM may generate a broad set of candidates to be filtered (or re-ranked) using weak supervision, this way increasing the overall system's robustness.\\[1ex]

{\small\noindent {\bfseries Acknowledgements:}
This work has received funding from the European Union’s Horizon Europe research and innovation programme under grant agreement No.\ 101070014 (OpenWebSearch.EU, \url{https://doi.org/10.3030/101070014}).}

%%
%% The acknowledgments section is defined using the "acks" environment
%% (and NOT an unnumbered section). This ensures the proper
%% identification of the section in the article metadata, and the
%% consistent spelling of the heading.
%%\begin{acks}
%%\end{acks}

%%
%% The next two lines define the bibliography style to be used, and
%% the bibliography file.
\bibliographystyle{ACM-Reference-Format}
\balance
\bibliography{main}

%%% -*-BibTeX-*-
%%% Do NOT edit. File created by BibTeX with style
%%% ACM-Reference-Format-Journals [18-Jan-2012].

\begin{thebibliography}{33}

%%% ====================================================================
%%% NOTE TO THE USER: you can override these defaults by providing
%%% customized versions of any of these macros before the \bibliography
%%% command.  Each of them MUST provide its own final punctuation,
%%% except for \shownote{} and \showURL{}.  The latter two
%%% do not use final punctuation, in order to avoid confusing it with
%%% the Web address.
%%%
%%% To suppress output of a particular field, define its macro to expand
%%% to an empty string, or better, \unskip, like this:
%%%
%%% \newcommand{\showURL}[1]{\unskip}   % LaTeX syntax
%%%
%%% \def \showURL #1{\unskip}           % plain TeX syntax
%%%
%%% ====================================================================

\ifx \showCODEN    \undefined \def \showCODEN     #1{\unskip}     \fi
\ifx \showISBNx    \undefined \def \showISBNx     #1{\unskip}     \fi
\ifx \showISBNxiii \undefined \def \showISBNxiii  #1{\unskip}     \fi
\ifx \showISSN     \undefined \def \showISSN      #1{\unskip}     \fi
\ifx \showLCCN     \undefined \def \showLCCN      #1{\unskip}     \fi
\ifx \shownote     \undefined \def \shownote      #1{#1}          \fi
\ifx \showarticletitle \undefined \def \showarticletitle #1{#1}   \fi
\ifx \showURL      \undefined \def \showURL       {\relax}        \fi
% The following commands are used for tagged output and should be
% invisible to TeX
\providecommand\bibfield[2]{#2}
\providecommand\bibinfo[2]{#2}
\providecommand\natexlab[1]{#1}
\providecommand\showeprint[2][]{arXiv:#2}

\bibitem[Alexander et~al\mbox{.}(2022)]%
        {alexander2022orcas}
\bibfield{author}{\bibinfo{person}{Daria Alexander}, \bibinfo{person}{Wojciech Kusa}, {and} \bibinfo{person}{Arjen P.~de Vries}.} \bibinfo{year}{2022}\natexlab{}.
\newblock \showarticletitle{ORCAS-I: queries annotated with intent using weak supervision}. In \bibinfo{booktitle}{\emph{Proceedings of the 45th International ACM SIGIR Conference on Research and Development in Information Retrieval}}. \bibinfo{pages}{3057--3066}.
\newblock


\bibitem[Bolotova et~al\mbox{.}(2022)]%
        {bolotova2022non}
\bibfield{author}{\bibinfo{person}{Valeriia Bolotova}, \bibinfo{person}{Vladislav Blinov}, \bibinfo{person}{Falk Scholer}, \bibinfo{person}{W~Bruce Croft}, {and} \bibinfo{person}{Mark Sanderson}.} \bibinfo{year}{2022}\natexlab{}.
\newblock \showarticletitle{A non-factoid question-answering taxonomy}. In \bibinfo{booktitle}{\emph{Proceedings of the 45th International ACM SIGIR Conference on Research and Development in Information Retrieval}}. \bibinfo{pages}{1196--1207}.
\newblock


\bibitem[Broder(2002)]%
        {broder2002taxonomy}
\bibfield{author}{\bibinfo{person}{Andrei Broder}.} \bibinfo{year}{2002}\natexlab{}.
\newblock \showarticletitle{A taxonomy of web search}. In \bibinfo{booktitle}{\emph{ACM Sigir forum}}, Vol.~\bibinfo{volume}{36}. ACM New York, NY, USA, \bibinfo{pages}{3--10}.
\newblock


\bibitem[Chandrakala et~al\mbox{.}(2024)]%
        {chandrakala2024intent}
\bibfield{author}{\bibinfo{person}{CB Chandrakala}, \bibinfo{person}{Rohit Bhardwaj}, {and} \bibinfo{person}{Chetana Pujari}.} \bibinfo{year}{2024}\natexlab{}.
\newblock \showarticletitle{An intent recognition pipeline for conversational AI}.
\newblock \bibinfo{journal}{\emph{International Journal of Information Technology}} \bibinfo{volume}{16}, \bibinfo{number}{2} (\bibinfo{year}{2024}), \bibinfo{pages}{731--743}.
\newblock


\bibitem[Craswell et~al\mbox{.}(2020)]%
        {craswell2020orcas}
\bibfield{author}{\bibinfo{person}{Nick Craswell}, \bibinfo{person}{Daniel Campos}, \bibinfo{person}{Bhaskar Mitra}, \bibinfo{person}{Emine Yilmaz}, {and} \bibinfo{person}{Bodo Billerbeck}.} \bibinfo{year}{2020}\natexlab{}.
\newblock \showarticletitle{Orcas: 18 million clicked query-document pairs for analyzing search}. In \bibinfo{booktitle}{\emph{Proceedings of the 29th ACM International Conference on Information \& Knowledge Management}}. \bibinfo{pages}{2983--2989}.
\newblock


\bibitem[Dubey et~al\mbox{.}(2024)]%
        {dubey2024llama}
\bibfield{author}{\bibinfo{person}{Abhimanyu Dubey}, \bibinfo{person}{Abhinav Jauhri}, \bibinfo{person}{Abhinav Pandey}, \bibinfo{person}{Abhishek Kadian}, \bibinfo{person}{Ahmad Al-Dahle}, \bibinfo{person}{Aiesha Letman}, \bibinfo{person}{Akhil Mathur}, \bibinfo{person}{Alan Schelten}, \bibinfo{person}{Amy Yang}, \bibinfo{person}{Angela Fan}, {et~al\mbox{.}}} \bibinfo{year}{2024}\natexlab{}.
\newblock \showarticletitle{The llama 3 herd of models}.
\newblock \bibinfo{journal}{\emph{arXiv preprint arXiv:2407.21783}} (\bibinfo{year}{2024}).
\newblock


\bibitem[Dubiel et~al\mbox{.}(2024)]%
        {dubiel2024device}
\bibfield{author}{\bibinfo{person}{Mateusz Dubiel}, \bibinfo{person}{Yasmine Barghouti}, \bibinfo{person}{Kristina Kudryavtseva}, {and} \bibinfo{person}{Luis~A Leiva}.} \bibinfo{year}{2024}\natexlab{}.
\newblock \showarticletitle{On-device query intent prediction with lightweight LLMs to support ubiquitous conversations}.
\newblock \bibinfo{journal}{\emph{Scientific Reports}} \bibinfo{volume}{14}, \bibinfo{number}{1} (\bibinfo{year}{2024}), \bibinfo{pages}{12731}.
\newblock


\bibitem[Figueroa(2015)]%
        {figueroa2015exploring}
\bibfield{author}{\bibinfo{person}{Alejandro Figueroa}.} \bibinfo{year}{2015}\natexlab{}.
\newblock \showarticletitle{Exploring effective features for recognizing the user intent behind web queries}.
\newblock \bibinfo{journal}{\emph{Computers in Industry}}  \bibinfo{volume}{68} (\bibinfo{year}{2015}), \bibinfo{pages}{162--169}.
\newblock


\bibitem[Han et~al\mbox{.}(2022)]%
        {han2022prompt}
\bibfield{author}{\bibinfo{person}{X Han}, \bibinfo{person}{W Zhao}, \bibinfo{person}{N Ding}, \bibinfo{person}{Z Liu}, {and} \bibinfo{person}{M~Sun Ptr}.} \bibinfo{year}{2022}\natexlab{}.
\newblock \showarticletitle{Prompt tuning with rules for text classification., 2022, 3}.
\newblock \bibinfo{journal}{\emph{DOI: https://doi. org/10.1016/j. aiopen}}  \bibinfo{volume}{3} (\bibinfo{year}{2022}), \bibinfo{pages}{182--192}.
\newblock


\bibitem[Hemied and Gabbar(2024)]%
        {hemied2024re}
\bibfield{author}{\bibinfo{person}{Omar~S Hemied} {and} \bibinfo{person}{Hossam~A Gabbar}.} \bibinfo{year}{2024}\natexlab{}.
\newblock \showarticletitle{RE-LLaMA: Large Language Models for Hydrogen Deployment: A Domain-Specific Approach in Renewable Energy}. In \bibinfo{booktitle}{\emph{2024 8th International Symposium on Innovative Approaches in Smart Technologies (ISAS)}}. IEEE, \bibinfo{pages}{1--4}.
\newblock


\bibitem[Jansen et~al\mbox{.}(2008)]%
        {jansen2008determining}
\bibfield{author}{\bibinfo{person}{Bernard~J Jansen}, \bibinfo{person}{Danielle~L Booth}, {and} \bibinfo{person}{Amanda Spink}.} \bibinfo{year}{2008}\natexlab{}.
\newblock \showarticletitle{Determining the informational, navigational, and transactional intent of Web queries}.
\newblock \bibinfo{journal}{\emph{Information Processing \& Management}} \bibinfo{volume}{44}, \bibinfo{number}{3} (\bibinfo{year}{2008}), \bibinfo{pages}{1251--1266}.
\newblock


\bibitem[Jansen et~al\mbox{.}(2007)]%
        {jansen2007web}
\bibfield{author}{\bibinfo{person}{Bernard~J Jansen}, \bibinfo{person}{Amanda Spink}, {and} \bibinfo{person}{Sherry Koshman}.} \bibinfo{year}{2007}\natexlab{}.
\newblock \showarticletitle{Web searcher interaction with the Dogpile. com metasearch engine}.
\newblock \bibinfo{journal}{\emph{Journal of the American Society for Information Science and Technology}} \bibinfo{volume}{58}, \bibinfo{number}{5} (\bibinfo{year}{2007}), \bibinfo{pages}{744--755}.
\newblock


\bibitem[Jin et~al\mbox{.}(2024)]%
        {jin2024comprehensive}
\bibfield{author}{\bibinfo{person}{Hanlei Jin}, \bibinfo{person}{Yang Zhang}, \bibinfo{person}{Dan Meng}, \bibinfo{person}{Jun Wang}, {and} \bibinfo{person}{Jinghua Tan}.} \bibinfo{year}{2024}\natexlab{}.
\newblock \showarticletitle{A comprehensive survey on process-oriented automatic text summarization with exploration of llm-based methods}.
\newblock \bibinfo{journal}{\emph{arXiv preprint arXiv:2403.02901}} (\bibinfo{year}{2024}).
\newblock


\bibitem[Kathuria et~al\mbox{.}(2010)]%
        {kathuria2010classifying}
\bibfield{author}{\bibinfo{person}{Ashish Kathuria}, \bibinfo{person}{Bernard~J Jansen}, \bibinfo{person}{Carolyn Hafernik}, {and} \bibinfo{person}{Amanda Spink}.} \bibinfo{year}{2010}\natexlab{}.
\newblock \showarticletitle{Classifying the user intent of web queries using k-means clustering}.
\newblock \bibinfo{journal}{\emph{Internet Research}} \bibinfo{volume}{20}, \bibinfo{number}{5} (\bibinfo{year}{2010}), \bibinfo{pages}{563--581}.
\newblock


\bibitem[Li et~al\mbox{.}(2024)]%
        {li2024pre}
\bibfield{author}{\bibinfo{person}{Junyi Li}, \bibinfo{person}{Tianyi Tang}, \bibinfo{person}{Wayne~Xin Zhao}, \bibinfo{person}{Jian-Yun Nie}, {and} \bibinfo{person}{Ji-Rong Wen}.} \bibinfo{year}{2024}\natexlab{}.
\newblock \showarticletitle{Pre-trained language models for text generation: A survey}.
\newblock \bibinfo{journal}{\emph{Comput. Surveys}} \bibinfo{volume}{56}, \bibinfo{number}{9} (\bibinfo{year}{2024}), \bibinfo{pages}{1--39}.
\newblock


\bibitem[Liu et~al\mbox{.}(2023)]%
        {liu2023we}
\bibfield{author}{\bibinfo{person}{Alisa Liu}, \bibinfo{person}{Zhaofeng Wu}, \bibinfo{person}{Julian Michael}, \bibinfo{person}{Alane Suhr}, \bibinfo{person}{Peter West}, \bibinfo{person}{Alexander Koller}, \bibinfo{person}{Swabha Swayamdipta}, \bibinfo{person}{Noah~A Smith}, {and} \bibinfo{person}{Yejin Choi}.} \bibinfo{year}{2023}\natexlab{}.
\newblock \showarticletitle{We're afraid language models aren't modeling ambiguity}.
\newblock \bibinfo{journal}{\emph{arXiv preprint arXiv:2304.14399}} (\bibinfo{year}{2023}).
\newblock


\bibitem[Liu et~al\mbox{.}(2022)]%
        {liu2022few}
\bibfield{author}{\bibinfo{person}{Haokun Liu}, \bibinfo{person}{Derek Tam}, \bibinfo{person}{Mohammed Muqeeth}, \bibinfo{person}{Jay Mohta}, \bibinfo{person}{Tenghao Huang}, \bibinfo{person}{Mohit Bansal}, {and} \bibinfo{person}{Colin~A Raffel}.} \bibinfo{year}{2022}\natexlab{}.
\newblock \showarticletitle{Few-shot parameter-efficient fine-tuning is better and cheaper than in-context learning}.
\newblock \bibinfo{journal}{\emph{Advances in Neural Information Processing Systems}}  \bibinfo{volume}{35} (\bibinfo{year}{2022}), \bibinfo{pages}{1950--1965}.
\newblock


\bibitem[Liu et~al\mbox{.}(2021)]%
        {liu2021makes}
\bibfield{author}{\bibinfo{person}{Jiachang Liu}, \bibinfo{person}{Dinghan Shen}, \bibinfo{person}{Yizhe Zhang}, \bibinfo{person}{Bill Dolan}, \bibinfo{person}{Lawrence Carin}, {and} \bibinfo{person}{Weizhu Chen}.} \bibinfo{year}{2021}\natexlab{}.
\newblock \showarticletitle{What Makes Good In-Context Examples for GPT-$3 $?}
\newblock \bibinfo{journal}{\emph{arXiv preprint arXiv:2101.06804}} (\bibinfo{year}{2021}).
\newblock


\bibitem[Liu and Shi(2024)]%
        {liu2024poliprompt}
\bibfield{author}{\bibinfo{person}{Menglin Liu} {and} \bibinfo{person}{Ge Shi}.} \bibinfo{year}{2024}\natexlab{}.
\newblock \showarticletitle{PoliPrompt: A High-Performance Cost-Effective LLM-Based Text Classification Framework for Political Science}.
\newblock \bibinfo{journal}{\emph{arXiv preprint arXiv:2409.01466}} (\bibinfo{year}{2024}).
\newblock


\bibitem[Mann et~al\mbox{.}(2020)]%
        {brown2020language}
\bibfield{author}{\bibinfo{person}{Ben Mann}, \bibinfo{person}{N Ryder}, \bibinfo{person}{M Subbiah}, \bibinfo{person}{J Kaplan}, \bibinfo{person}{P Dhariwal}, \bibinfo{person}{A Neelakantan}, \bibinfo{person}{P Shyam}, \bibinfo{person}{G Sastry}, \bibinfo{person}{A Askell}, \bibinfo{person}{S Agarwal}, {et~al\mbox{.}}} \bibinfo{year}{2020}\natexlab{}.
\newblock \showarticletitle{Language models are few-shot learners}.
\newblock \bibinfo{journal}{\emph{arXiv preprint arXiv:2005.14165}}  \bibinfo{volume}{1} (\bibinfo{year}{2020}).
\newblock


\bibitem[Mohasseb et~al\mbox{.}(2019)]%
        {mohasseb2019customised}
\bibfield{author}{\bibinfo{person}{Alaa Mohasseb}, \bibinfo{person}{Mohamed Bader-El-Den}, {and} \bibinfo{person}{Mihaela Cocea}.} \bibinfo{year}{2019}\natexlab{}.
\newblock \showarticletitle{A customised grammar framework for query classification}.
\newblock \bibinfo{journal}{\emph{Expert Systems with Applications}}  \bibinfo{volume}{135} (\bibinfo{year}{2019}), \bibinfo{pages}{164--180}.
\newblock


\bibitem[Morishita et~al\mbox{.}(2025)]%
        {morishita2025enhancing}
\bibfield{author}{\bibinfo{person}{Terufumi Morishita}, \bibinfo{person}{Gaku Morio}, \bibinfo{person}{Atsuki Yamaguchi}, {and} \bibinfo{person}{Yasuhiro Sogawa}.} \bibinfo{year}{2025}\natexlab{}.
\newblock \showarticletitle{Enhancing Reasoning Capabilities of LLMs via Principled Synthetic Logic Corpus}.
\newblock \bibinfo{journal}{\emph{Advances in Neural Information Processing Systems}}  \bibinfo{volume}{37} (\bibinfo{year}{2025}), \bibinfo{pages}{73572--73604}.
\newblock


\bibitem[Mosbach et~al\mbox{.}(2023)]%
        {mosbach2023few}
\bibfield{author}{\bibinfo{person}{Marius Mosbach}, \bibinfo{person}{Tiago Pimentel}, \bibinfo{person}{Shauli Ravfogel}, \bibinfo{person}{Dietrich Klakow}, {and} \bibinfo{person}{Yanai Elazar}.} \bibinfo{year}{2023}\natexlab{}.
\newblock \showarticletitle{Few-shot fine-tuning vs. in-context learning: A fair comparison and evaluation}.
\newblock \bibinfo{journal}{\emph{arXiv preprint arXiv:2305.16938}} (\bibinfo{year}{2023}).
\newblock


\bibitem[Pass et~al\mbox{.}(2006)]%
        {pass2006picture}
\bibfield{author}{\bibinfo{person}{Greg Pass}, \bibinfo{person}{Abdur Chowdhury}, {and} \bibinfo{person}{Cayley Torgeson}.} \bibinfo{year}{2006}\natexlab{}.
\newblock \showarticletitle{A picture of search}. In \bibinfo{booktitle}{\emph{Proceedings of the 1st international conference on Scalable information systems}}. \bibinfo{pages}{1--es}.
\newblock


\bibitem[Rose and Levinson(2004)]%
        {rose2004understanding}
\bibfield{author}{\bibinfo{person}{Daniel~E Rose} {and} \bibinfo{person}{Danny Levinson}.} \bibinfo{year}{2004}\natexlab{}.
\newblock \showarticletitle{Understanding user goals in web search}. In \bibinfo{booktitle}{\emph{Proceedings of the 13th international conference on World Wide Web}}. \bibinfo{pages}{13--19}.
\newblock


\bibitem[Rouzegar and Makrehchi(2024)]%
        {rouzegar2024enhancing}
\bibfield{author}{\bibinfo{person}{Hamidreza Rouzegar} {and} \bibinfo{person}{Masoud Makrehchi}.} \bibinfo{year}{2024}\natexlab{}.
\newblock \showarticletitle{Enhancing Text Classification through LLM-Driven Active Learning and Human Annotation}.
\newblock \bibinfo{journal}{\emph{arXiv preprint arXiv:2406.12114}} (\bibinfo{year}{2024}).
\newblock


\bibitem[Russell et~al\mbox{.}(2009)]%
        {russell2009task}
\bibfield{author}{\bibinfo{person}{D Russell} {et~al\mbox{.}}} \bibinfo{year}{2009}\natexlab{}.
\newblock \showarticletitle{Task Behaviors During Web Search: The Difficulty of Assigning Labels‖, Proceedings of the 42nd Hawaii International Conference on System Sciences (HICSS)}.
\newblock  (\bibinfo{year}{2009}).
\newblock


\bibitem[Sun et~al\mbox{.}(2023)]%
        {sun2023llm}
\bibfield{author}{\bibinfo{person}{Xiaofei Sun}, \bibinfo{person}{Xiaoya Li}, \bibinfo{person}{Jiwei Li}, \bibinfo{person}{Fei Wu}, \bibinfo{person}{Shangwei Guo}, \bibinfo{person}{Tianwei Zhang}, {and} \bibinfo{person}{Guoyin Wang}.} \bibinfo{year}{2023}\natexlab{}.
\newblock \showarticletitle{Text classification via large language models}.
\newblock \bibinfo{journal}{\emph{arXiv preprint arXiv:2305.08377}} (\bibinfo{year}{2023}).
\newblock


\bibitem[Tan et~al\mbox{.}(2023)]%
        {tan2023can}
\bibfield{author}{\bibinfo{person}{Yiming Tan}, \bibinfo{person}{Dehai Min}, \bibinfo{person}{Yu Li}, \bibinfo{person}{Wenbo Li}, \bibinfo{person}{Nan Hu}, \bibinfo{person}{Yongrui Chen}, {and} \bibinfo{person}{Guilin Qi}.} \bibinfo{year}{2023}\natexlab{}.
\newblock \showarticletitle{Can ChatGPT replace traditional KBQA models? An in-depth analysis of the question answering performance of the GPT LLM family}. In \bibinfo{booktitle}{\emph{International Semantic Web Conference}}. Springer, \bibinfo{pages}{348--367}.
\newblock


\bibitem[Wei et~al\mbox{.}(2022)]%
        {wei2022chain}
\bibfield{author}{\bibinfo{person}{Jason Wei}, \bibinfo{person}{Xuezhi Wang}, \bibinfo{person}{Dale Schuurmans}, \bibinfo{person}{Maarten Bosma}, \bibinfo{person}{Fei Xia}, \bibinfo{person}{Ed Chi}, \bibinfo{person}{Quoc~V Le}, \bibinfo{person}{Denny Zhou}, {et~al\mbox{.}}} \bibinfo{year}{2022}\natexlab{}.
\newblock \showarticletitle{Chain-of-thought prompting elicits reasoning in large language models}.
\newblock \bibinfo{journal}{\emph{Advances in neural information processing systems}}  \bibinfo{volume}{35} (\bibinfo{year}{2022}), \bibinfo{pages}{24824--24837}.
\newblock


\bibitem[Xu et~al\mbox{.}(2024)]%
        {xu2024llms}
\bibfield{author}{\bibinfo{person}{Hanzi Xu}, \bibinfo{person}{Renze Lou}, \bibinfo{person}{Jiangshu Du}, \bibinfo{person}{Vahid Mahzoon}, \bibinfo{person}{Elmira Talebianaraki}, \bibinfo{person}{Zhuoan Zhou}, \bibinfo{person}{Elizabeth Garrison}, \bibinfo{person}{Slobodan Vucetic}, {and} \bibinfo{person}{Wenpeng Yin}.} \bibinfo{year}{2024}\natexlab{}.
\newblock \showarticletitle{LLMs' Classification Performance is Overclaimed}.
\newblock \bibinfo{journal}{\emph{arXiv preprint arXiv:2406.16203}} (\bibinfo{year}{2024}).
\newblock


\bibitem[Yadav et~al\mbox{.}(2024)]%
        {yadav2024pag}
\bibfield{author}{\bibinfo{person}{Vikas Yadav}, \bibinfo{person}{Zheng Tang}, {and} \bibinfo{person}{Vijay Srinivasan}.} \bibinfo{year}{2024}\natexlab{}.
\newblock \showarticletitle{Pag-llm: Paraphrase and aggregate with large language models for minimizing intent classification errors}. In \bibinfo{booktitle}{\emph{Proceedings of the 47th international ACM SIGIR conference on research and development in information retrieval}}. \bibinfo{pages}{2569--2573}.
\newblock


\bibitem[Zhang et~al\mbox{.}(2024)]%
        {zhang2024clamber}
\bibfield{author}{\bibinfo{person}{Tong Zhang}, \bibinfo{person}{Peixin Qin}, \bibinfo{person}{Yang Deng}, \bibinfo{person}{Chen Huang}, \bibinfo{person}{Wenqiang Lei}, \bibinfo{person}{Junhong Liu}, \bibinfo{person}{Dingnan Jin}, \bibinfo{person}{Hongru Liang}, {and} \bibinfo{person}{Tat-Seng Chua}.} \bibinfo{year}{2024}\natexlab{}.
\newblock \showarticletitle{CLAMBER: A Benchmark of Identifying and Clarifying Ambiguous Information Needs in Large Language Models}.
\newblock \bibinfo{journal}{\emph{arXiv preprint arXiv:2405.12063}} (\bibinfo{year}{2024}).
\newblock


\end{thebibliography}

\end{document}